\documentclass[prl, twocolumn,nofootinbib, letterpaper, noraggedfooter,nobibnotes ]{revtex4}

\usepackage{amssymb,amsmath,amsfonts}
\usepackage{graphics,dcolumn,bm,fleqn,epic,eepic,float}
\usepackage{graphicx} % Include figure files
\usepackage{bm} % bold math
\usepackage{subfigure}
\usepackage{xcolor}
\usepackage{xr}
\begin{document}

%\title{Complex processes and the context sensitivity of Entropy}{}
\title{Equivalence of information production and generalized entropies in complex processes}{}

\author{Rudolf Hanel$^{a,b}$, Stefan Thurner$^{a,b,c}$}

%\email[Corresponding Author. E-mail: ]{ stefan.thurner@meduniwien.ac.at}

\affiliation{$^a$Section for Science of Complex Systems, CeMSIIS, Medical University of Vienna, Spitalgasse 23, A-1090, Austria}
\affiliation{$^b$Complexity Science Hub Vienna, Josefst\"adter Strasse 39, A-1080 Vienna, Austria}
\affiliation{$^c$Santa Fe Institute, 1399 Hyde Park Road, Santa Fe, NM 85701, USA}

\date{Version \today}

\begin{abstract}
Complex systems that are 
%in particular those 
characterized by strong correlations and fat-tailed distribution functions 
%in their dynamical variables, 
have been argued to be incompatible within the framework of Boltzmann-Gibbs entropy. 
As an alternative, so-called generalized entropies were proposed and intensively studied. 
Here we show %on information theoretic grounds 
that this incompatibility is a misconception. 
For a broad class of processes, Boltzmann entropy --the log multiplicity-- remains the valid entropy concept, however, for non-i.i.d., non-multinomial, and non-ergodic processes, Boltzmann entropy is not of Shannon form, $-k\sum_i p_i \log p_i$. 
The correct form of Boltzmann entropy can be shown to be identical with generalized entropies.
We derive this result for all processes that can be mapped reversibly to adjoint representations where processes are i.i.d..  In these representations the information production is given by the Shannon entropy. We proof that over the original sampling space this yields functionals that are identical to generalized entropies. The problem of constructing adequate context-sensitive entropy functionals therefore can be translated into the much simpler problem of finding adjoint representations. The method provides a comprehensive framework for a statistical physics of strongly correlated systems and complex processes. 
\end{abstract}
%\pacs{02.70.Rr, 02.50.Cw, 05.10.Gg}% PACS

\maketitle

%%%%%%%%%%%%%%%%%%%%%%%%%%%%%%%%%%%%%%%%%%%%%%%
%%%%%%%%%%%%%%%%%%%%%%%%%%%%%%%%%%%%%%%%%%%%%%%
\section{Introduction}
%%%%%%%%%%%%%%%%%%%%%%%%%%%%%%%%%%%%%%%%%%%%%%%
%%%%%%%%%%%%%%%%%%%%%%%%%%%%%%%%%%%%%%%%%%%%%%%

To know the information content of a process, a system, a  source, a signal, or a sequence, 
one uses entropy to quantify it. If systems or processes are independent identically distributed (i.i.d.), 
ergodic and stationary in their probabilities, it is known what to do: one uses the expression \cite{Shannon1948},  
\begin{equation}\label{eq:shannon_entropy}
H(p)=-\sum_{i=1}^W p_i\log p_i  \,,
\end{equation} 
where $i=1,2,\cdots,W$ are the states the system can take and $p_i$ is the probability to observe them.
So-called Shannon entropy is given by $S(p)=k H(p)$, where $k>0$ is a constant that specifies the units of entropy, 
in statistical physics $k=k_B$ is the Boltzmann constant; if we measure information in bits per symbol, $k=1/\log 2$.
If the system or process of interest is not i.i.d., ergodic, or in stationary equilibrium, then it becomes less clear what 
to do in order to obtain its correct information content. In principle there are two conceptually very different paths to solve the problem. 
\\

{\bf The first way} is to look at the {\em information production} of the process or system. Consider a 
process, $X$, that emits signals $x(T)$ of length $T$ with $x(T)=x_T x_{T-1}\cdots x_1$. Such 
an ordered list/sequence of elements $x(T)$ is often referred to as a $T$-tuple.
Every symbol, $x_t$, in the signal is an element of a fixed sample space or an alphabet, ${\cal A}$, 
that contains all possible states $x_t$ can take. 
For example, if you think of $X$ as a text-producing author, states are the letters from the 
English alphabet,  ${\cal A^{\rm letters}}$, 
or in a ``binary alphabet'',  ${\cal A}^{0,1}$ of zeros and ones, once the text is stored on a computer. 
In non-i.i.d. systems, symbols within sequences will in general be correlated in one way or another.
Those correlations --that may extend over many different scales in the system-- carry information about the system. 
Using the marginal distribution functions of the occurrences of states (letters) in Eq. (\ref{eq:shannon_entropy}) 
will then certainly not provide the correct information content of the process. 
However, if we know the probability distribution, $p_T$, to observe 
entire sequences, $x(T)$, we compute the information production of process, $X$, as 
\begin{equation}\label{eq:info_prod}
I(X)=\lim_{T\to\infty} \frac{1}{T} k H(p_T)\, ,
\end{equation}
see also SI Text 2 for details. 
This changes the perspective from individual symbols, events, or states to entire sequences, or paths.
From an information theoretic point of view, $I(X)$ measures the average number of bits required to 
reversibly encode samples of $X$ into bit-streams that can be sent through information channels 
and measure information production in \textit{bits per emitted symbol}, i.e., $k=1/\log(2)$. 
Note that $k H(p_T)$ measures bits per $T$-tuple, i.e. per path-segment-of-length-$T$. Therefore, by using
$k_T=k/T$, one measures again bits per emitted symbol of the original alphabet.  
However, the number of all possible $T$-tuples --the size of the new ``alphabet''-- is enormous if $T$ is large. 
If the sample space contains $A=|{\cal A}|$ symbols, the size of the alphabet for all paths is $A^T$.
The sum over all states and knowing their probabilities, $p_T$, will in general be impossible. 

The reason why $I(X)$ is the true information production rate of process $X$, is because there exist representations 
of $X$ in terms of other processes, $Y$, such that sequences, $x$, presented in the symbols from the initial alphabet, ${\cal A}_{0}$, 
can be rewritten into sequences, $y$, using other symbols from a much larger alphabet. The definition of information production, $I(X)$, asymptotically uses the largest alphabet containing each possible sequence as a unique symbol, which --as a consequence-- are statistically independent. 
In other words, all structure (correlations) gets absorbed into new symbols belonging to an {\em extended alphabet}. Only, the definition of information production uses the largest possible alphabet; the alphabet 
of $T$-tuples over the original alphabet and $T\to\infty$.

However, typically, there exist much smaller alphabets that can capture all structures of a process or system. 
I.e. we can find a shorter representation, $Y$, of the process, $X$. 
% and translate signals emitted by $X$ to the symbols of the extended alphabet used by the respective representation $Y$. 
We say $Y$ is {\em adjoined} to $X$
(see subsection \ref{sec:adjointprocessspace} for details) if $Y$s  symbols are uncorrelated and, thus, $H$ measures the correct information content. In general, the symbols, $z$, in the extended alphabet, encode for a number, $\ell$, symbols in the original alphabet. For instance, if $z$ is a symbol in the alphabet of $T$-tuples, then $T=\ell(z)$, $z$ consisting exactly of 
$T$ letters of the original alphabet. 
The average length, $\bar\ell=\langle\ell\rangle=\sum_z f_z \ell(z)$, 
of original symbols emitted per symbol in the extended alphabet increases with the size of the extended alphabet. 
$f_z=p_z(Y)$ is the distribution function of letters $z$. %that uniquely represents $X$ in terms of representation $Y$.
Consequently, the unit of information, $k$, \textit{adapts} to the "complexity" $\ell(z)$ and $k\to k/\bar\ell$. 
%In alphabets of  $T$-tuples used in the definition of information production we have $\bar\ell=T$.
%{\gr One may note here that} it is the essence of science in general and ML {\gr (maximum likelihood)} and AI techniques in particular to capture these correlations in data in efficient ways. 

For an example of how one can encode information about correlations of  a process $X$ on all relevant scales imagine an initial alphabet of Latin letters and extend it to a series of extended alphabets: one that contains syllables in addition to letters, one that adds word-fragments, one that includes words, one with frequent word combinations, one with phrases, and so on. This sequence of alphabets is nested in the sense that they contain each other, $\mathcal{A}_{0}\subset\mathcal{A}_{1}\subset\mathcal{A}_{2}\subset \cdots \subset\mathcal{A}_{n} \cdots$. With any one of these alphabets, say $\mathcal{A}_{n}$, one can sample ``text'' from it  by using the associated marginal distributions, $p_\lambda(n)$, of its elements, $\lambda$. With increasing $n$, the resulting artificial text samples will more and more resemble the  English text body from which the marginal distributions $p_\lambda(n)$ were derived; see SI Text 1 for the examples given by C. Shannon. 
We can find particular sequences of alphabets such that each alphabet, $\mathcal{A}_{n+1}$, contains exactly one more symbol  than $\mathcal{A}_{n}$, by applying reversible substitutions of symbols, which we refer to as ``parsing rules''. 
For details, see SI Text 3.
\\

{\bf The second way} is to capture the correlations and structures in non-i.i.d. systems / processes in a generalized functional form of  the entropy, which --typically--  looks more complicated than Eq. (\ref{eq:shannon_entropy}). Generalized entropy functionals are usually expressed in terms of marginal distributions in the ``original alphabet''. These {\em generalized entropies} have been extensively studied for several decades \cite{genent1,genent2,genent3, HTMGM3, HTgendriven2018} from different angles, generally for systems with strong or long-range correlations \cite{cproc1,cproc2,cproc3}, that are non-ergodic, internally constrained \cite{constr1,constr2,constr3}, or for systems out of equilibrium \cite{nequil1,nequil2,nequil3}.

For non-i.i.d. systems or processes it is essential to specify the context in which the term entropy is used, whether one talks about information theory,  thermodynamics, or  the maximum entropy principle (MEP) \cite{3faces}. While thermodynamic aspects of such systems, especially the existence of well defined thermodynamic potentials (and as a consequence, temperature) is heavily debated, there is wide consensus that {\em entropy production} (the physical analogue of information production) remains a valid concept, also for these systems. Here we will not focus on thermodynamic  aspects of entropy, but on the original context envisioned by Boltzmann: its power to predict a particular macro state from knowing the number of micro states (multiplicity) corresponding to it. This allows one to predict typical distribution functions and derive functional relations between macro state variables (expectation values of respective observables). This view that is tightly related to the MEP is not restricted to physics and is not limited to i.i.d. processes. For specific cases the respective MEP functionals, the generalized entropy and cross entropy, have been explicitly constructed \cite{HTMGM3, HTgendriven2018}. In particular, if multiplicity and probabilities are multiplicatively separable in the assymptotic limit, \cite{HTMGM3}, a clear definition of cross entropy is possible also for non-i.i.d. systems.

While the Boltzmann entropy concept remains untouched (log multiplicity) its functional form, i.e., the generalized entropy functional, depends on the context of the process, $X$, and the process class, $\Phi$, to which it belongs to. Some examples for different process classes include i.i.d. processes, exchangeable and polynomial mixture processes \cite{TRS_ExchProc}, Polya processes \cite{HCMT2017}, sample space reducing processes \cite{HTgendriven2018}, and processes describing structure forming systems \cite{Jan_StructureFormingSyst}.

The idea behind generalized entropies is to quantify entropy as the logarithm of the number of micro states (multiplicity) of process, $X$. It is based on the marginal distribution function, $g_i=p_i(X)$, of symbols $i$ from a sample space (here the original alphabet is $\mathcal{A}_0$) that compose a functional, $\tilde S(g)$, such that  it captures the structural information in $X$. Similarly, one obtains generalized expressions for cross entropy and information divergence. Those functionals effectively capture arbitrarily complicated relations between symbols of the sample space (original alphabet) in terms of marginal symbol frequencies, $g$, of the process $X$. 
Generalized entropies (that fulfil the first Shannon-Khinchin axiom) do not explicitly depend on system parameters that identify a process within a process class or other details. It is obvious that in general, constructing such a functional may be complicated and has not been achieved convincingly, except for a few exceptions, e.g. \cite{HCMT2017,HTgendriven2018}.  
As we will see, one can reconstruct (or at least approximate) such functionals from data -- at least in principle, 
since there exist fundamental limits to reconstructing generative grammars, 
from data on the basis of statistical inference alone, \cite{Chomsky1,Chomsky2}, a fact also captured in Chaitin's incompleteness theorem, \cite{Chaitin}. In other words, the question of whether some data, a particular sequence, $x$, contains regular structures that can be used to compress it, may become undecidable.
\\ 

\begin{figure}[t]
\begin{center}
		\includegraphics[width=0.9\columnwidth]{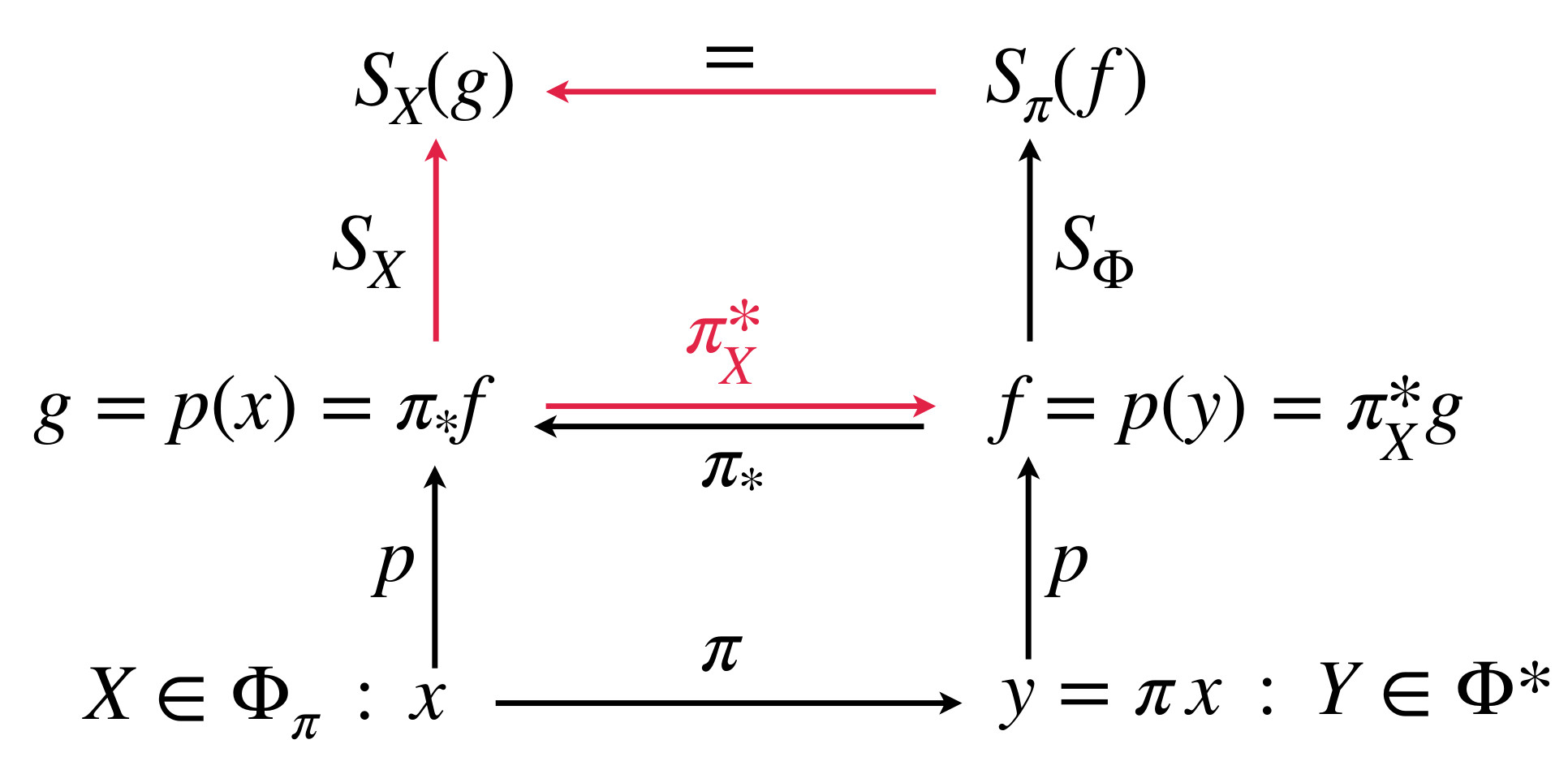}
\end{center}
	\caption{	
	 Diagram of the relations between of distribution functions and entropies over the
	 sample space and adjoint samples space. We consider a process, $X$ over alphabet $\mathcal{A}$ with adjoint
	 process $Y=\pi X$ over the adjoint alphabet $\mathcal{A}^*$. $Y$ is i.i.d. and therefore 
	 fully characterized by the marginal distribution of letters it samples from, i.e. asymptotically  
	 data $y=\pi x$ is fully characterized by the relative frequency distribution function $f=p(y)$.	 
	 $\Phi^*$ is the set of all i.i.d. processes over $\mathcal{A}^*$. 
	 Therefore $\Phi_\pi = \pi^{-1} \Phi^*$ is the class of processes that naturally generalize $X\in \Phi_\pi$. 
	 $f$ can be projected to the marginal distribution function $g=p(x)=\pi_* p(y)=\pi_* f$. 
	 Conversely, for a particular process, $X$, we can lift the distribution function $g$ to the associated
	 adjoint distribution and get $\pi^*_X g=\pi^*_X p(x)=p(y)=f$. Since $Y$ is i.i.d. over the 
	 adjoint sample space one can measure information production simply by using Shannon entropy
	 with the adequate Boltzmann factor $k_Y$ (adapted to the distribution function $f$). The
	 commutative diagram therefore defines the process specific generalized entropy $S_X$ over 
	 the sample space of the process class $\Phi_\pi=\pi^{-1}\Phi^*$. 
}	 	
	\label{fig:diagram}
\end{figure}

Here we show that the two approaches, information production and the generalized entropy functionals can be mapped to one another, meaning that they are the same. The diagram in Fig. (\ref{fig:diagram})  schematically shows the basic idea: We first use the method of parsing rules to construct an adjoint representation of a given process $X$, to a process $Y=\pi X$. Here $\pi$ is a map that encodes all structures in $X$, such that process $Y$ in its new extended alphabet is i.i.d. and is therefore fully described by their marginal distribution function, $f_z=p_{z}(Y)$, where $z$ is again a letter from the extended alphabet.
Consequently, the Shannon information measure with the appropriate unit of information is adequate (first way). 
Next, we project the marginal distributions from the adjoint representation, $f_z$, to the original alphabet and, in a last step, 
we identify  the ``pull-back'' information measures that takes distribution functions over the original alphabet, with the corresponding generalized information measure (second way).

For the proof we use the \textit{minimal description length} (MDL) (see also SI Text 2), the length of the shortest encoding that fully represents the data. In this context,  we shall see that information production is tightly related to the notion of Kolmogorov complexity \cite{Kolmogorov,Solomonov,Chaitin}; For a brief discussion, see SI Text 4. We explicitly demonstrate the method in an example for the class of sample space reducing (SSR) processes \cite{HCMT2017} that provide simple, analytically tracktable models for driven dissipative systems. Their generalized entropy is exactly known for arbitrary driving rates \cite{HCMT2017,HTgendriven2018}. 

The purpose of the paper is to show that indeed, $S_X(X)=k_YH(Y)$. The proof is given in the following section.

\section{Results}

The key tool used in the following are {\em parsing rules}, simple substitution rules that allow us to reversibly re-code (possibly correlated) data streams into new symbol streams that no-longer carry structures; see SI Text 3. The simplest parsing rule template is denoted by $[r\ s\to m]$ and means that two symbols $r$ and $s$ that appear frequently together (the pair $rs$ is over-represented with respect to their marginal probabilities of appearance) are substituted with a new symbol $m$. In the following we will associate $m$ also with the symbol index, meaning that it is the $m$-th symbol in an alphabet.  
This is the elementary parsing rule template. The particular choice of a set of parsing rule templates, we call {\em the parsing rule}.
We speak of a \textit{relevant} set of parsing rule templates if
(i) one can extract the full information content of a process, $X$, asymptotically, solely by using parsing rules from the set of templates, and (ii) if omitting one template from the set does not allow one to do so.
In the following we focus on processes for which the elementary parsing rule template forms a relevant set. However, the arguments  presented extend naturally to more general sets of parsing rule templates; see SI Text 3. 

\subsection{Constructing adjoint process spaces}\label{sec:adjointprocessspace}

To construct adjoint representations one can proceed as follows. Suppose $X$ is a process that emits symbols $i$ drawn from the alphabet $\mathcal{A}_0=\{1,2,\cdots,W\}$, where $W$ is the number of symbols. $X$ generates data streams, $x(t)=x_{t}x_{t-1}\cdots x_{1}$, where every $x_t$ is one of the available symbols in $\mathcal{A}_0$,
which contains $W$ elements. Consider now two letters $r_1$ and $s_1$ such that the pair $r_1s_1\equiv x_{\tau}x_{\tau-1}$ is over-expressed in the data $x$ or has been identified to contain relevant information by any other method of inference, then we can rewrite the pair $r_1s_1$ by a new letter $m_1=W+1$, which will become the first letter extending alphabet $\mathcal{A}_0$ to $\mathcal{A}_1=\mathcal{A}_0\cup \{W+1\}$, 
with the parsing rule $\pi_1=[r_1\ s_1\to W+1]$. 

We can iterate and produce parsing rules 
$\pi_n=[r_n\ s_n \to W+n]$, with letter indices $r_n<W+n$ and $s_n<W+n$.
Where $\pi_n$ maps data over the alphabet $\mathcal{A}_{n-1}$ to data over 
$\mathcal{A}_{n}=\mathcal{A}_{n-1}\cup \{W+n\}$.
Note that the parsing rules  $\pi_n$ can be uniquely inverted, i.e. we can expand data over $\mathcal{A}_{n}$ to data over $\mathcal{A}_{n-1}$ using the inverse map $\pi_n^{-1}= [ W+n\ \to\ r_n\ s_n]$.
In other words the inverse parsing rules can be thought of being part of a "generative grammar", \cite{Chomsky1}.
We therefore can construct a sequence of maps $\pi(n)=\pi_n\pi_{n-1}\cdots\pi_1$ such that data $x$
can be mapped to representations $y_n=\pi(n)x$. At every \textit{parsing level}, $n$, we get a corresponding distribution function of the re-coded data, $p_z(y_n)$, with a letter index $z=1,2,\cdots , W+n$.

The Kraft and McMillan theorems \cite{Kraft, McMillan}  tells us that if all that we know about a process are its marginal relative frequencies, $g_i$, at which symbols $i$ occur, there exists a shortest reversibly encoding of the data, $x$, of characteristic length, $L^{\rm min}$, the {\em minimal description length} (MDL) that gives the theoretically achievable minimal length of $x$ (in units of bits). 
$L_{\rm min}$ is a lower bound for the true MDL, $L(x)$, that can only be attained asymptotically. The theorems  state that
\begin{equation}
k H(g)=\lim_{t\to\infty} \frac{1}{t}L(x(t))\,,
\end{equation}
with $k=1/\log(b)$, where $b$ is the basis  in which information is measured. For bits we typically have $b=2$.
$k H(g)$ is the MDL in bits per symbol and $L^{\rm min}\,=\,t k H(g)$ is the minimal number of bits required to encode messages of length $t$.

For data $x(t)$ of length $t$ we find a sequence of representations $y_n(t)=\pi(n)x(t)$. 
Suppose now that for every $t$ we find a parsing level $n^*(t)$ such that 
$y_{n^*(t)}(t)$ is a representation of the data $x(t)$ that cannot be distinguished from an i.i.d. process (which we indicate here by a $*$). It follows that  $y_{n^*(t)}(t)$ is entirely determined by its marginal distribution
of letters $p_z(y_{n^*(t)}(t))$ and obtain asymptotically 
$k H(p(y_{n^*(t)}(t)))\simeq \frac{1}{|y_{n^*(t)}(t)|}L^{\rm min}(y_{n^*(t)}(t))$,
where $\simeq$ means asymptotically identical. 
With $|.|$ we indicate the length of the sequence of letters in numbers of letters of the underlying alphabet. For instance we have $|x(t)|=t$ and $|y_{n+1}(t)|\leq |y_{n}(t)|$. Then, we can asymptotically measure the information production of the process $X$ to be 
\begin{equation}
\frac{1}{t}L^{\rm min}(x(t))\simeq\frac{|y_{n^*(t)}(t)|}{|x(t)|} k H(p(y_{n^*(t)}(t)))\,.
\end{equation}
As discussed above, the ``complexity'' of a symbol $\ell(z)$ is the number of letters in the original alphabet it codes for. As a consequence we can compute the average symbol complexity for data $y_{n^*(t)}(t)$ to be given by 
$\langle \ell \rangle_{Y_{n^*(t)}} \simeq |x(t)|/|y_{n^*(t)}(t)|$.
As a consequence, we get for the adjoint process $Y=\lim_{t\to\infty} Y_{n^*(t)}$ that 
$\langle \ell \rangle_{Y}=\lim_{t\to\infty} \langle \ell \rangle_{Y_{n^*(t)}}$. 
The adequate unit of information, $k_{Y}$, for measuring information production, $k_{Y} H(p(Y))$, therefore is given by 
\begin{equation}
k_{Y}(x)=\frac{k}{\langle \ell \rangle_{Y}}\,.
\label{kY}
\end{equation}
For simplicity, suppose there is a maximal $n^*$ that holds for all $t$. Since we assumed that $y_{n^*}$ is already indistinguishable from an i.i.d. process, applying another parsing rule would only compress the data without changing its MDL. This means, $k_{Y_{n*+1}} H(p(Y_{n^*+1}))= k_{Y_{n^*}} H(p(Y_{n^*}))$. If on the contrary we look at a parsing level, $n$, where the adjoint process is not yet i.i.d., then we can find a parsing rule, $\pi_{n+1}$, such that $k_{Y_{n+1}} H(p(Y_{n+1})) < k_{Y_{n}} H(p(Y_{n}))$. Additional knowledge always reduces the attainable information production rate. 

In principle, for any finite amount of data $x(t)$ one can construct the optimal map, $\pi(n)$, for the process, 
$X$, by minimizing over all possible sequences of parsing rules, at any fixed parsing level $n$, 
if only we know the relevant set of parsing rules to consider and this set is finite.
Then we can in principle find $n^*(t)$ by finding that $n$ such that no further reduction of the minimal 
description length is possible by applying any more parsing rules.
In practice, an extensive search over all possible sequences of parsing rules is of course not feasible, even if 
the set of parsing rule templates only consists of the elementary template,   
and algorithms for inferring adjoint representations of data need to turn to different means of optimization. 
For theoretical considerations we may, however, assume that for a given finite relevant set of parsing rule templates and any finite $t$ we can find the optimal map $\pi$ (or one of several possible optimal maps if the map is not unique) or at least a map reasonably close to optimal, since the number of possible maps we would have to evaluate remains finite, too. Intuitively it is clear however, given an unknown process $X$ for which we cannot pre determine the respective relevant set of parsing rule templates, that one typically can no longer decide whether a map $\pi$ is optimal or not.
 
However, Given an optimal $\pi$, the adjoint i.i.d. process, $Y$, is fully characterized by its marginal distribution, $f=p(Y)$, over symbols in the adjoint alphabet, $\mathcal{A}^*\equiv \mathcal{A}_{n^*}$, and the information production, $I(X)=k_{Y} H(f)$, is given by the Shannon entropy of $Y$. On the adjoint process space $\Phi^*$ we can use the measures of Shannon entropy, cross-entropy, and Kullback-Leibler information divergence, given that we use the appropriate unit of information, $k_Y$ of Eq. (\ref{kY}).
Since $Y$ is i.i.d. over  $\mathcal{A}^*$, the adjoint space naturally belongs to the family, $\Phi^*$, of \textit{all} i.i.d. processes over this alphabet. Further, any $Y'$ in $\Phi^*$ is fully characterized by its marginal distribution, $f'$, and the pair $(f',\pi)$, determines the process $X'=\pi^{-1}Y'$. Hence, the process class, $\Phi_\pi$, that naturally generalizes a process, $X$, with adjoint i.i.d. process, $Y=\pi X$, is  given by $\Phi_\pi=\pi^{-1}\Phi^*$.  

This construction completes the first part of the proof that establishes that we can essentially measure information production of a process as the Shannon entropy of the adjoint process. This entropy, however uses the marginal distributions over the adjoint alphabet as arguments and can therefore not be identified directly with the generalized entropies that use the marginal distributions over the original alphabet as arguments. In the next step we will pull the information measures over the adjoint message space back to to the original message space.

\subsection{Information measures over extended alphabets}
 
Suppose we have a process $X$ with an adjoint process $Y=\pi X$.
For data, $x$, and its adjoint sequence, $y=y_n$, we obtain two histograms, $h_i(x)$, of symbols $i\in\mathcal{A}_0$ and, $h_z(y), $of symbols $z\in\mathcal{A}_n$, respectively. The associated relative frequency distributions are given by $g=p(x)=h(x)/|x|$ and $f=p(y)=h(y)/|y|$. Further, every symbol, $z$, represents a number of $\ell(z)$ symbols, $\pi^{-1} z$, in the original alphabet with $\pi=\pi(n)$. 
We define $\bar h_i(z)=h_i(\pi^{-1} z)$ as the histogram of letters $i\in\mathcal{A}_0$ that are parsed together into the symbol $z\in\mathcal{A}_n$. For $z\leq W$ where $W=|\mathcal{A}_0|$, we have, $\bar h_i(z)=\delta_{iz}$ and $h_i(x)=\sum_{z\in\mathcal{A}_n} \bar h_i(z)h_z(y)$ needs to hold for all $i\in\mathcal{A}_0$. This provides us with constraints,
\begin{equation}
 h_i(x)=\sum_{z\in\mathcal{A}_n}\bar h_i(z)h_z(y)\, ,
\label{equ:matchingconstr1}
\end{equation}
that we need in the next section. 
As a consequence, we have 
\begin{equation}
g_i(x)=\frac{|y|}{|x|}\sum_{z\in\mathcal{A}^*}\bar h_i(z)f_z(y)\,.
\end{equation}
We drop the arguments $(x)$ and $(y)$ (or $(y_n)$) from now on and distinguish histograms by their index $i$ (over $\mathcal{A}_0$) and $z$ (over the adjoint alphabet $\mathcal{A}^*=\mathcal{A}_n$). 
Let $\langle \ell\rangle_{f} = \sum_{z\in\mathcal{A}^*} f_z \ell(z) $ (note that in this notation we identify $\langle \ell\rangle_{f}\equiv \langle \ell\rangle_{Y}$) and 
$\langle \bar h_i\rangle_{f}  = \sum_{z\in\mathcal{A}^*} f_z \bar h_i (z) $ be the expectation values under the distribution $f$, then, by construction, $|x|=\sum_{z\in\mathcal{A}^*} h_z \ell(z)$,   
$|y|=\sum_{z\in\mathcal{A}^*} h_z$, and $k_Y(x)=k\langle \ell\rangle_{f}^{-1}$. 
This means that we can write the constraints that link distributions $g$ over $\mathcal{A}_0$, Eq. (\ref{equ:matchingconstr1}), with distributions $f$ over $\mathcal{A}^*$ in the following way
\begin{equation}
0\ =\ C_i(g|f)\ \equiv\  \langle \ell\rangle_{f}^{-1}  \langle \bar h_i \rangle_{f}\ -\ g_i \,.
\label{equ:matchingconstr2}
\end{equation}

As mentioned before, the process class, $\Phi_\pi$, that $X$ belongs to, is completely determined by the map $\pi$ and the process $X$, by the pair $(f,\pi)$, see Fig. 2.
Therefore, we can identify the entropy of $X$ with
\begin{equation}
S_{\pi}(f) \equiv  k_Y H(f)\,  ,
\label{eq:entropy}
\end{equation}
with the process-specific Boltzmann factor, $k_Y\,\equiv\, k/(\langle \ell \rangle_f)$. For processes, $X$ and $X'$, and with $f'=p(\pi X')$, the cross-entropy and the information divergence are 
\begin{equation}
\begin{array}{lcl}
S^{\rm cross}_{\pi}(f||f') &\equiv& k_Y H_{\rm cross}(f||f') \\
& &\\
D_{\pi}(f'||f) &\equiv& k_{Y} D_{\rm KL}(f'||f) \, . 
\end{array}
\label{eq:crossanddkl}
\end{equation}
In the special case where $X$ is already an i.i.d. process, no features can be extracted from the data and $n=0$, $\pi=\pi(0)=id$, and $\ell(z)=1$, for all $z=1\cdots W$. Consequently, $S_{\pi}=k H$, $S^{ \rm cross }_{\pi}=k H_{ \rm cross }$, and $D_{\pi}=k D_{\rm KL}$ (Kullback-Leibler divergence), as required. 

\subsection{Pulling back entropies to the original alphabet}

In the next step on can construct entropy functionals over the original alphabet, $\mathcal{A}_0$, by lifting a distribution function, $g'$, on $\mathcal{A}_0$ to a distribution function, $f'$, over $\mathcal{A}^*$ by assuming that $f=p(\pi X)$ is the true distribution function of the process, $Y=\pi X$. We proceed by minimizing the information divergence, $D_{\pi}(f'||f)$, with respect to $f'$. More precisely, we minimize the functional $\psi(f',\alpha,\eta)$ given by
\begin{equation}
%D_{\pi}(f'||f)-\alpha\left(|f'| -1\right)
%-\sum_{i\in\mathcal{A}_0}\eta_i\left(\frac{\langle \bar h_i\rangle_{f'}}{\langle \ell\rangle_{f'}}-g'_i\right)\ ,
D_{\pi}(f'||f)-\alpha\left(|f'| -1\right)
-\sum_{i\in\mathcal{A}_0}\eta_i\ C_i(g'|f')\ ,
\label{eq:entropy2}
\end{equation}
with Lagrange multipliers, $\alpha$ and $\eta_i$, that normalize $f'$ and guarantee the  constraints from Eq. (\ref{equ:matchingconstr2}). Solving the variational principle $\delta\psi=0$ estimates $f'$ at the maximum that is compatible with $g'$. We identify this maximizer as $f^{(g')}\equiv f'$ and obtain $\alpha=1/\langle \ell\rangle_{f}$ and
\begin{equation}
f^{(g')}_z = f_z\, e^{ \left(  
D_{\pi} (f^{(g')}||f) + 
\sum\limits_{ i\in\mathcal{A}_0 } \frac{ \eta_i }{ \langle \ell\rangle_{f} } 
\left( \frac{\bar h_i(z) }{ \ell(z) }-g'_i   \right) 
\right)\ell(z) } \, ,
\label{eq:distr}
\end{equation}
which has to be solved self consistently. If $f$ already meets all matching constraints with $g'$, i.e. if $g'=g$ with $g=p(X)$, then we have  
$D_{\pi}(f^{(g)}||f)=0$ and $f^{(g)}=f$.

This means that one can lift marginal distributions, $g'$, on $\mathcal{A}_0$ to distributions, $f^{(g')}$, on $\mathcal{A}_n$ with respect to a particular process, $X$. As a consequence one can pull the entropy, cross-entropy, and divergence back from distributions, $f$, over the adjoint sample space, $\mathcal{A}_n$, to distributions, $g$, over the initial alphabet, $\mathcal{A}_0$, with respect to a particular process, $X\in\Phi_\pi$. In particular, one can define the projection operator, $\pi_*$, through $g=p(\pi^{-1}\,y) \equiv \pi_*p(y)=\pi_* f$ and the operator, $\pi^*_X$, that lifts distributions, $g$, over alphabet, $\mathcal{A}_0$, to  distributions $f$ over the extended alphabet, $\mathcal{A}^*$, through
\begin{equation}
\pi^*_X g\ \equiv\ {\rm minarg}_{\{f'\,|\,g=\pi_*f'\}}\ D_{\pi}(f'||p(\pi\, X)) \,,
\label{eq:lift}
\end{equation}
with respect to the process $X$, i.e., with respect to the distribution function, $p(\pi\, X)$, of the i.i.d. process, $\pi\, X$, over the adjoint alphabet. The lift operator gives us the maximizers, $f^{(g)}=\pi^*_X\,g$. We find that $id=\pi_*\pi^*_X$ and identify
\begin{equation}
\begin{array}{lcl}
\bar S_X(g) &\equiv& S_{\pi}(\pi^*_X g)\\
& &\\
\bar S^{\rm cross}_{X}(g'||g) &\equiv& S^{\rm cross}_{\pi}(\pi^*_X g'||\pi^*_X g)\\
& &\\
\bar D_X(g'||g) &\equiv & D_{\pi}(\pi^*_X g'||\pi^*_X g)\, ,
\end{array}
\label{eq:entcrossanddkl}
\end{equation}
where, typically, $g=p(X)$, and call them the {\em pull-back} entropy, cross-entropy, and information 
divergence of the process $X$. 
Note, that while $S_{\pi}$, $S^{\rm cross}_{\pi}$, and $D_{\pi}$ are universal on 
the entire class of processes $\Phi_\pi$ pulling the information measures back to to marginal distributions $g$ over $\mathcal{A}_0$ gives information measures that are specific to a particular process $X\in\Phi_\pi$. 
% NOTATION IST SUPER SO!!! HOECHSTENS VERWIRREND DASS DER STERN OBEN FUER IID VERWENDET WURDE

\subsection{Generalized entropies over initial alphabets}

The final question is how the pull-back measures $S_{X}$, $S^{\rm cross}_{X}$, and  $D_{X}$ are related to generalized entropy functionals as derived for example in \cite{HCMT2017,HTgendriven2018}. There, functionals were derived to obtain the most likely histogram, $h$, observed in a given process after $t$ observations (for large $t$). Even for non i.i.d. processes, often the probability, $P(h|\theta)$, to observe the particular histogram, $h$, for $t=\sum_{i\in\Omega}h_i$ observations factorizes $P(h|\theta)=M(h) G(h |\theta)$ into a {\em multiplicity}, $M(h)$, and a probability term of the sequences, $G(h |\theta)$. $\theta$ is a set of parameters that determines the process, $X$ -- it defines and parametrizes the process class. Whenever such a factorization is possible, one can show that a generalized maximum entropy principle exists. Using the Boltzmann definition of entropy, the logarithm of multiplicity, $\tilde S=\log M/t$, and defining $\tilde S^{\rm cross}=-\log G/t$, and a generalized information divergence as 
%\begin{equation}
$\tilde D(g|\theta)=- t^{-1}\log P(h|\theta)$, %\,,
%\label{eq:gendivergence}
%\end{equation}
where $g=h/t$, the standard relations $\tilde D=\tilde S^{\rm cross}-\tilde S$ remain valid \cite{HTMGM3}. 
%To define entropy as a function that only depends on probabilities (first Shannon-Khinchin axiom), one requires $\tilde S$ to be independent of $\theta$. This is sometimes possible but typically it is only partially possible. For SSR processes $\tilde S$, for instance, is completely independent of one set of system parameters specifying the proportions of the SSR states, but it depends implicitly on coupling constants to a driving process, \cite{HTgendriven2018}. In the example one can still decouples $\tilde S$  from $\theta$ by including the distribution function of the coupling process into $\tilde S$ too. That is, $\tilde S$ then depends only on the joint probabilities of the process and the coupling process together but no longer on the coupling constants. 

We now try to find a map $\pi$ such that data from all processes $X(\theta)$ decorrelate over the adjoint alphabet $\mathcal{A}^*$. $X(\theta)$ forms a sub-class of processes in $\Phi_{\pi}$ with distribution functions, $g(\theta)=p(X(\theta))$,  and adjoint distribution functions, $f(\theta)=p(\pi X(\theta))$.
If $\bar \ell(\theta) = \sum_z f_z(\theta)\ell(z)$, $y'=\pi x'$ and $t=|x'|= |y'|\bar\ell(\theta)$, we can asymptotically re-express $P(h(x)|\theta)$ in terms of the multinomially distributed histograms $h'$ over the adjoint alphabet,
\begin{equation}
P(h'|f(\theta))={\frac{t}{\bar\ell(\theta)} \choose{h'}} \prod_{z\in\mathcal{A}^*}\, f_i(\theta)^{h'_z}\,.
\end{equation}
If we collect all those adjoint sequences, $y'$, with $\bar\ell=|x'|/|y'|$ that fulfil the matching constraint of Eq. (\ref{equ:matchingconstr2}) between $g'=p(x')=h(x')/t$ and $f(\theta)$ and if we call this set $\Gamma(x',\theta)$, we get 
\begin{equation}
P(h(x')|\theta)=\sum_{y'\in\Gamma(x',\theta)} P(h(y')|f(\theta))\,.
\label{eq:lala}
\end{equation}
For large $t$ the sum on the right hand side of Eq. (\ref{eq:lala}) can be approximated by the maximal term contributing to the sum and
\begin{equation}
P(h(x')|\theta)\simeq \max_{y'\in \Gamma(x',\theta)} P(h(y')|f(\theta))\,.
\label{eq:lala2}
\end{equation}
Taking logs and multiplying both sides with $-1/t$ we obtain
\begin{equation}
\begin{array}{lcl}
D(g'|\theta) &\simeq& \min\limits_{y'\in \Gamma( x',\theta)} k_{Y(\theta)} D_{\rm KL}(p(y')|f(\theta)) \\
&=&\min\limits_{y'\in \Gamma(x',\theta)} D_{\pi}(p(y')||f(\theta))\\
&=& D_{X(\theta)}(g'|g(\theta))\,,
\end{array}
\label{eq:lalala}
\end{equation}
where $\simeq$ means asymptotically identical for large $t$; from the second to the third line we used the definition of the lift operator from Eq. (\ref{eq:lift}) and  $D_X$ from Eq. (\ref{eq:entcrossanddkl}). In other words, we have shown that in the limit $t\to\infty$ the generalized information divergence $D(g'|\theta)$ is identical to the pull-back divergence $D_{X(\theta)}(g'|g(\theta))$ as a functional. As a consequence we can use that $D_X=S^{\rm cross}_X-S_X$ and identify $S_X$ with the generalized entropy and $S^{\rm cross}_X$ with the generalized cross entropy. We see that the generalized entropy, $\tilde S$, for the processes family, $X(\theta)$, is given by $\tilde S(g)=S_{X(\theta)}(g)$. 

% DAS VIELLEICHT ALS  SI TEXT XXX There are several ways to read the last line. If the generalized entropy fulfils the first Shannon-Khinchin axiom (SK1) (entropy only depends on probabilities and nothing else), then $S_{X(\theta)}(g)$ is invariant on any choices of $\theta$. But one can also read it in the way that  the generalized $\tilde S(g)$ will not fulfil the first SK1 and $\tilde S$ explicitly depends on the system parameters, $\theta$. This may then (thinking of \cite{HTgendriven2018}) make us inclined to search for a larger picture, where the process $X$ is a sub process of another process containing $X$, for which the respective $\tilde S$ decouples from $\theta$. However, this may be $\tilde S(g)=S_{X(\theta)}(g)$ whenever the family $X(\theta)$ fulfils tSK1 and  $S_{X(\theta)}(g)$ does not depend on $\theta$. When this is not the case the pull-back entropy  $S_{X(\theta)}(g)$ still is the appropriate generalized entropy but violates SK1. 

We conclude that %Boltzmann entropy still rules. We have shown here, 
for all process classes (at least those that accept elementary parsing rules) there exist notions of entropy, $S_{X}$, cross-entropy, $S^{\rm cross}_{X}$, and divergence, $D_{X}$, that behave in the usual way, namely, $D_X=S^{\rm cross}_X-S_X$. This means that for such processes (at least asymptotically) the probability to observe a particular histogram, $P\simeq \exp (-t D_X)$, factorizes into a multiplicity term, $M\simeq \exp (t S_X)$, associated with entropy, and a sequence probability term, $G\simeq \exp (- t S^{\rm cross}_X)$, associated with cross-entropy. In other words, we have shown that Boltzmann's entropy, the logarithm of  multiplicity, remains the correct estimator of information production. For complex systems the multiplicity will differ from a multinomial coefficient in arbitrarily complicated ways that might even depend explicitly on system parameters $\theta$, which would mean a violation of SK1. In other words, the Boltzmann pull-back entropy functional typically will be more complicated than the Shannon entropy functional and can even violate SK1 -- yet they are the appropriate generalizations of entropy. We also learned that beyond
such parametric families of generalized entropies, i.e. beyond the pull-back measures, we find the standard notions of entropy, cross-entropy, and divergence  present in the adjoint alphabet, where the only thing that is not universal about the information measures is the process-specific Boltzmann constant, $k_Y$, that needs to be used. 
%REDUNDANT It is therefore not a question of changing our concept of entropy, which basically remains what Boltzmann already had in his mind, log multiplicity, but about the functional form that entropy takes when we apply the entropy concept to different process classes where the multiplicity of processes paths sharing the same probability may vary wildly between  process classes in general and from a multinomial multiplicity factor (the multinomial coefficient) in particular.

%{\re XXXXXXXXXXX}

\subsection{Example: SSR processes}

We now demonstrate explicitly that the generalized entropy that --according to the previous section-- is identified with the pull-back entropy, $S_{X}$, %of the respective process class, 
indeed does measure information production. We do that by considering slowly driven sample space reducing (SSR) processes, $X$, for which the generalized entropy functional is exactly known \cite{HCMT2017}. SSR processes are models of driven non-equilibrium systems. They are characterized by the fact that as the process unfolds the number of states accessible to the process reduces when no driving is present \cite{CMHT2015}. In its simplest form, the process relaxes to a ground state from which it has to be restarted. One can think of the process as a ball bouncing down a staircase with random jump sizes. The ball can only  jump to steps lower than the last step it visited. Once it reaches the bottom of the staircase one lifts the ball to the top of the staircase (driving), and kicks it down the staircase again. The stairs represent (energy) states, the lowest being $1$, the highest is $W$. The process exhibits path-dependence in the relaxation part, the current through the system breaks detailed balance between states. 

The micro-states, $x$, of the SSR process are sequences of states with elements $x_n\ \in\  \{1,2, \cdots, W\}\ \equiv\ \Omega$. The transition probabilities between states $j\ to i$ are
\begin{equation}
q(i|j)=\Theta(j-i) q_i/Q_{j-1}\ +\ \delta_{j,1}q_i\ \, 
\label{eq:SSRtransprob}
\end{equation}
where the first term on the right hand side describes the relaxing part of the SSR process (transitions  only happen from higher $i$ to lower states, $j$, i.e., when $j<i$) with prior distribution $q_i$ and cumulative distribution, $Q_j=\sum_{i=1}^j q_i$. $\Theta$ is the Heavyside function. The second term captures the (slow) driving of the process. Slow here means that the system is only driven once the SSR process reaches its lowest position $i=1$. SSR processes are Markovian since transition probabilities depend only on the current position and it is ergodic since after the relaxation process the system is reset to any state with probability, $q_i$.  

To understand the statistics of the process we are interested in the distribution of visits to the individual states. We define the macro-state %$\mu$, 
to be the histogram, $h_i$, of visits of $X$ to state, $i$. It is possible to compute $S_{\rm SSR}(p)=\frac1t\ \log M(k)$, where the multiplicity, $M(k)$, is the number of different sequences, $x$, of length $t$ with the same histogram $h$. One finds \cite{HCMT2017} 
\begin{equation}
S_{\rm SSR}(p)=-\sum_{i=2}^W  p_i\log \frac{p_i}{p_1}+(p_1-p_i)\log \frac{p_1-p_i}{p_1} \, ,
\label{eq:SSRent}
\end{equation} 
where $p_i=h_i/t$ are the relative frequencies of observing a state $i$.
Note that this is the Boltzmann entropy of the system, yet it is not of Shannon form since it is derived from a Markov, and not an independent sampling process. Similarly, one finds the cross-entropy 
\begin{equation}
S^{\rm cross}_{\rm SSR}(p|q)=-\sum_{i=1}^W p_i\log q_i + \sum_{i=2}^W p_i\log Q_{i-1}\quad,
\label{eq:SSRcross}
\end{equation} 
and by maximizing $\psi=S_{\rm SSR} -S^{\rm cross}_{\rm SSR}$, ({\rm negative} information divergence $D$), one obtains the characteristic Zipf distribution of the slowly driven SSR process \cite{CMHT2015}
\begin{equation}
p_i=p_1\frac{q_i}{Q_{i}}  \, .
\label{eq:SSRfin}
\end{equation} 
For the special case of $q_i=1/W$ for all $i$, the Zipf distribution, $p_i \sim \frac{1}{i}$, is obvious,
since $Q_i=i/W$ and $q_i/Q_i=1/i$. It continues to hold for ``well-behaved'' $q_i$, 
\cite{Robustness_CMHT2016}. For instance, if $q_i\propto i^\alpha$ for $\alpha>-1$, then $Q_i\propto i^{1+\alpha}$ and again $q_i/Q_i\propto 1/i$.

In the next step we will use a simple example of a slowly driven SSR process 
in order to demonstrate how using extended alphabets works and 
how the respective generalized entropy functional is the adequate measure of 
information production. 

\subsubsection{Example of a small SSR system}

To demonstrate how a minimal adjoint alphabet for a slowly driven SSR process looks like, consider such a process over an initial alphabet of $W=4$ symbols (numbers) representing the four states, $i\in\mathcal{A}_0=\{1,2,3,4\}$. A SSR sequence in that alphabet might look like $x=421214321431212141\cdots$. Remember that 
the $q_i$ are normalized weights such that the probability to sample the state $ j<i$ conditional on the process being in state $i$ is given by $q_j/Q_i$, with $Q_i=\sum_{j=1}^i q_j$ and by $q_j$ if the system is in the 
ground state $i=1$ and the system gets driven. 
%{\re REMIND HERE WHAT THE $q_i$ s mean. SUM to 1???}
%THIS IS TOO COMPLICATED We know that the conditional entropy works for all Markov processes and therefore the adjoint alphabet of the conditional entropy for a $4$-state Markov processes consists of pairs of letters $\mathcal{A}_{n^*}=\mathcal{A}_0^2$, with $n^*=W^2$. SSR processes are a specific class of Markov processes and we therefore  expect the adjoint alphabet to capture SSR specific structure. In fact there is only one independent letter transition in a slowly driven SSR process. That is when the process gets restarted after having reached state $i=1$. That is, 
One can think of the adjoint SSR alphabet, $\mathcal{A}_{n^*}$, as the union of $\mathcal{A}_0$ with the set of new symbols that represent all possible strictly monotonic decreasing sequences on 
$\mathcal{A}_0$, i.e., $\mathcal{A}_{n^*}=\{1,2,3,4,5,6,7,8,9,10,11\}$, where the new symbol "5" represents the sequence $2\,1$, "6" stands for $3\,1$, "7" for $3\,2\,1$, "8" for $4\,1$, "9" for $4\,2\,1$, "10" for $4\,3\,1$, 
and "11" for $4\,3\,2\,1$. Since we have 7 new symbols, $n^*=7$ extending the alphabet of original symbols $\{1,2,3,4\}$ we have a total of $11$ symbols in the extended alphabet  $\mathcal{A}_{n^*}$.
The $7$ parsing rules producing the new symbols are given by 
\begin{equation}
\begin{array}{lcl}
\pi_1&=&[2\ 1\to 5]\\
\pi_2&=&[3\ 1\to 6]\\
\pi_3&=&[3\ 5\to 7]\\
\pi_4&=&[4\ 1\to 8]\\
\pi_5&=&[4\ 5\to 9]\\
\pi_6&=&[4\ 6\to 10]\\
\pi_7&=&[4\ 7\to 11]\,
\end{array}
\end{equation} 
the map $\pi=\pi_7\pi_6\pi_5\pi_4\pi_3\pi_2\pi_1$ can be constructed that maps between messages written in the initial and the adjoint alphabet. We therefore can rewrite our example $x=421214321431212141\cdots$ into
$\pi(1) x=4554354315541\cdots$, then $\pi(2) x=455435465541\cdots$, $\pi(3) x=45547465541\cdots$,
$\pi(4) x=4554746558\cdots$, $\pi(5) x=954746558\cdots$, $\pi(6) x=9547[10]558\cdots$, and finally 
$\pi(7) x=95[11][10]558\cdots$.
We now project a distribution function, $f$, on $\mathcal{A}_{n^*}$ to a distribution function, $g$, 
on $\mathcal{A}_{0}$. 
Note that all the new extended letters with index 5 to 11 represent subsequences that contain a $1$.
That is $\bar h_1(z)=\delta_{1z}+\sum_{s=5}^11\delta_{sz}$. The original letter 2 is part of the subsequences 
represented by the extended letters $5$ (the sequence $2\,1$), $7$ (the sequence $3\,2\,1$), $9$ (the sequence $4\,2\,1$), and $11$ (the sequence $4\,3\,2\,1$). That is $\bar h_2(z)=\delta_{1z}+\delta_{5z}+\delta_{7z}+\delta_{9z}+\delta_{11z}$. Similarly we can find $\bar h_i(z)$ for $i=3$ and $i=4$.
As a consequence the distribution functions 
$g_i$ of the message $x$ in original letters $i=1,\cdots,4$ and the distribution function 
$f_z$ of the adjoint message $\pi(7)x$ in extended letters $z=1,2,\cdots,11$ are given by
four equations:
\begin{equation}
\begin{array}{lcl}
g_1&=&\frac1Z(f_1+f_5+f_6+f_7+ \cdots\\
&&\qquad + f_8+f_9+f_{10}+f_{11})\\
g_2&=&\frac1Z(f_2+f_5+f_7+f_9+f_{11})\\
g_3&=&\frac1Z(f_3+f_6+f_7+f_{10}+f_{11})\\
g_4&=&\frac1Z(f_4+f_8+f_9+f_{10}+f_{11})\,,
\end{array}
\label{eq:example:projection}
\end{equation}
where $Z$ is a normalization constant such that $1=\sum_{i=1}^4 g_i$.
Note that after applying $\pi$ to a SSR process yields (asymptotically) that $f_2=f_3=f_4=0$. We can now express the asymptotic relative frequencies, i.e. the probabilities, $f_z$, in terms of the weights $q_i$ on the SSR states $i=1,2,3,4$, and get, $f_5=q_2$, $f_6=q_3q_1/(q_1+q_2)$, $f_7=q_3q_2/(q_1+q_2)$, $f_8=q_4q_1/(q_1+q_2+q_3)$, and so forth. Inserting the expressions for $f_z$ in Eq. (\ref{eq:example:projection})  one self-consistently obtains the marginal distribution on the original alphabet
\begin{equation}
g_i=\frac{1}{Z}\frac{q_i}{Q_i}\,,
\end{equation}    
as predicted from Eq. (\ref{eq:SSRfin}); -- 
note that if $q_i=1/W$ is uniform, then the solution $q_i/Q_i=1/i$ is exactly reproducing Zipf's law
and for a broad variety of choices for $q_i$ one obtains approximate Zipf laws.
That means that $f$ fulfils the matching constraints of Eq. (\ref{equ:matchingconstr2}) exactly and therefore also the lift, $\pi^*_X$, of the asymptotic marginal distribution function, $g$, to $f$ is exact and is given by $f=\pi^*_X g$. That is, we can see in this simple example how
the distribution function $g$ over the original alphabet can be predicted from knowing the distribution function $f$ of letters of the extended alphabet.

Since slowly driven SSR processes are actually Markov processes one can also proof that the respective 
generalized entropies are actually the adequate information measures in this context.
It is well known that the information production of a Markov process can be measured by the 
so-called conditional entropy, $S_{\rm cond}$. This is a functional depends on the 
probabilities $p^{(2)}=p_{ij}$, that a symbol $j$ follows symbol $i$ in the process. 
The SSR entropy on the other hand on the other hand depends on $p^{(1)}=p_{i}$ is the marginal distribution 
of the symbols, $i$. If $p^{(2)}$ is the maximizer of the conditional entropy, or more precisely, the minimizer of the conditional information divergence, and $p^{(1)}$ is the minimizer of the SSR information divergence, then both estimators of entropy, the conditional entropy and the SSR entropy, are identical, 
$S_{\rm cond}(p^{(2)})\equiv S_{\rm SSR}(p^{(1)})$, for all choices of the system parameters $q$.
For details of the computation, 
% Rudi : CHECK
see SI Text 5.
%{\bl see SI Text 5}.
%%%%%%%%%%%%%%
%$S_{\rm SSR}$ does not explicitly depend on the system parameters $q\equiv \theta$ and therefore fulfils SK1.

\section{Discussion}

We showed that by identifying the entropy of a process with its information production it is possible to consistently extend the fundamental notion of entropy in statistical physics --Boltzmann entropy-- to non-i.i.d. processes and processes that operate out of equilibrium. This is done by identifying isomorphisms that map entire process classes an adjoint representation where processes are i.i.d.. The sample space (or alphabet) of the adjoint process is typically larger than the sample space of the original process. The isomorphisms can be thought of concatenations of parsing rules that map strongly correlated segments in the original process to new symbols. Information production of the adjoint i.i.d. process is quantified by Shannon entropy. Pulling back the entropy measure in the adjoint space to the original sample space and comparing the resulting functional with the Boltzmann entropy (process-specific log multiplicity) establishes the asymptotic equivalence of the notion of generalized entropy and information production.

This provides a comprehensive image that consistently links information theory and the statistical physics of non i.i.d. processes in a   context-sensitive way that allows us to associate an entropy, a cross-entropy (representing the constraints of the maximum entropy principle), and an information divergence (or relative entropy) to complex processes. Context-sensitive means that the functional form of the entropy depends on the class of processes considered; the concept of entropy  itself, information production from the information theoretic perspective and Boltzmann entropy from the perspective of physics, remains invariant.

If an adjoint representation of one process is found, one can find the adjoint representations of an entire class of processes that all de-correlate in their representations over the same adjoint sample space. This means that there exists a natural way how processes implicitly define their own generalization to an entire process class. This is possible because the property of de-correlating over the same adjoint sample space implements an equivalence relation. This has important consequences since these equivalence classes of processes generalize the idea of an ensemble to non-i.i.d. processes which provides a concise way --grounded in first principles-- to extend the program of statistical physics to complex processes and their macro variables.

%The presented results also provide us with a venture point for constructing methods that efficiently infer adjoint process representations and thus the corresponding entropic measures of a process from data. However, it has to be noted that there are limits to the ability of statistical methods to detect structure (as e.g. Chomsky or Chaitin have noted) and statistical methods can in general only give us proxies of adjoint representations and their respective entropy measures.

Supported by Austrian Science Fund FWF Projects under I3073 and P34994 and Austrian Research Promotion Agency, FFG project under 873927.

%{\re BIBLIOGR. IS A MESS -- PLEASE FIX AND USE CORRECT TITLES NAMES, ETC. Maybe USE A BIB FILE}

\newpage

\section*{Supplementing Information}

\subsection{SI Text 1: Shannon's example of random texts from different alphabets}

Shannon develops his intuition underlying his definition of information production from examples he takes from language, i.e. how differently artificially generated text looks like, if one samples text using the English alphabet $\mathcal{A}_{\rm letter}$ and (1) only uses the marginal distribution $p(x_t)$, with $a=x_t$  being letters $a\in\mathcal{A}_{\rm letter}$, here is the example Shannon gives:
\\
"OCRO HLI RGWR NMIELWIS EU LL NBNESEBYA TH EEI ALHENHTTPA OOBTTVA NAH BRL",
\\
or if (2) one uses information on letter transition probabilities  $p(a'|a)$, for letter $a,a'\in\mathcal{A}_0$ in English texts, 
\\
"ON IE ANTSOUTINYS ARE T INCTORE ST BE S DEAMY ACHIN D ILONASIVE TUCOOWE AT TEASONARE FUSO TIZIN ANDY TOBE SEACE CTISBE",
\\
or if one (3)  switches to the word level with the alphabet  $\mathcal{A}_{\rm word}$ of English words and samples $p(x_t)$, with $w=x_t$  being words $w\in\mathcal{A}_{\rm word}$, 
\\
"REPRESENTING AND SPEEDILY IS AN GOOD APT OR COME CAN DIFFERENT NATURAL
HERE HE THE A IN CAME THE TOOF TO EXPERT GRAY COME TO FURNISHES
THE LINE MESSAGE HAD BE THESE",
\\
or if (4) one uses information on letter transition probabilities  $p(w'|w)$, for words $w,w'\in\mathcal{A}_{\rm word}$ in English texts,
\\
"THE HEAD AND IN FRONTAL ATTACK ON AN ENGLISH WRITER THAT THE CHARACTER
OF THIS POINT IS THEREFORE ANOTHER METHOD FOR THE LETTERS THAT
THE TIME OF WHO EVER TOLD THE PROBLEM FOR AN UNEXPECTED".

\subsection{SI Text 2: Minimal description length, i.i.d. processes, and compression}

Let us briefly look at how embedding messages into larger alphabets works for compressing i.i.d. sequences. The Kraft-McMillan theorem tells us that we can find a uniquely decodable prefix code over a code alphabet of length $r$ (typically binary $r=2$) for the states $i\in\Omega=\{1,\cdots,W\}$ with length of the codewords $\ell_i$ if and only if $\sum_{i=1}^W r^{-\ell_i}$. Let $h_i(x)$ be the number of times the symbol $i$ appears in the message $x$ and
$p_i=h_i/t$ is the marginal frequency distribution of states $i$ in the same message. 
If one chooses $\ell_i=\lceil \log(1/p_i)/\log r\rceil$, where $\lceil \log(1/p_i)\rceil$ is the natural number such that $\lceil y \rceil\ \geq\ y\ >\ \lceil y\rceil-1$, then those $\ell_i$ satisfy the Kraft-McMillan theorem and for the description length $L(x)$ of the process we get that $L=\sum_{i=1}^W h_i \lceil \log(1/p_i)/\log r\rceil$ and as a consequence we obtain
\begin{equation}
H(p)/\log r+1> \frac{L}{t} \geq H(p)/\log r\,
\end{equation}
where $H(p)=\sum_{i=1}^W p_i \log(1/p_i)$ is Shannon entropy.
Since we assume the process is i.i.d. 
then $H(p)/\log r$ is also the minimal description length (MDL) per symbol $i=1,\cdots,W$ that can be asymptotically achieved as the data volume, $t=|x|$, gets large. 

However, MDL typically cannot be fully obtained for a message written in its original alphabet. 
A binary process emitting only zeros and ones cannot be made any shorter by encoding single zeros and ones differently. The information theoretic way to show that $H(p)/\log r$ is the asymptotically obtainable lower limit of the MDL, is by considering extended alphabets, for instance $\Omega^2$ with new letters $i'=ij\in \Omega^2$  that are 2-tuples of the original letters. If the process is i.i.d. then we also know that for instance the probability of $p'_{i'}=p_ip_j$ and from the additivity of $H$ we obtain $H(p')=2H(p)$. If we transform the message of even length $t$, $x(t)=x_{t}x_{t-1}x_{t-2}\cdots x_1$, into a sequence of half the length written in 2-tuple letters 
$x'(t/2)=(x_{t}x_{t-1})(x_{t-2}x_{t-3})\cdots (x_2x_1)$, we have $t\to t'=t/2$ and we get that $H(p')/\log r+1> \frac{L}{t'} \geq H(p)/\log r$. As a consequence we get for the MDL $L$, in the 2-tuple alphabet $\Omega^2$, that $H(p)/\log r+1/2> \frac{L}{t} \geq H(p)/\log r$. Considering $2^n$-tuple letters $i'\in \Omega^{(2^n)}$ one gets for the effective minimal information rate that $H(p)/\log r+2^{-n}> \frac{L}{t} \geq H(p)/\log r$. In other words, by considering larger and larger alphabets one can rewrite i.i.d. messages into a code alphabet (something we will never do here) in such a way that asymptotically one finds that $\frac{L}{t} \simeq H(p)/\log r$,  asymptotically approaching the lower bound from above. The same asymptotic result can be obtained for i.i.d. processes by using parsing maps $\pi(n)=\pi_n\pi_{n-1}\cdots \pi_1$, similar to the way described in the main paper, only that we search for parses that reduce the description length without the requirement that the reduction in description length is higher then the one expected for a respective i.i.d. process. however, in this way parsing is used for pure data compression since i.i.d. data has no inherent features. 

\subsection{SI Text 3: Generative grammars, parsing rules and parsing rule templates}

Intuitively parsing rules are particular rules that tell us how to replace some symbols 
in a sequence, signal, or text, with other symbols. A parsing rule template characterizes not the particular rule but the way substitution rules are constructed.

Say for instance, you look for occurrences of the letter $b$, that we indicate as symbol $(b)$, following the letter $a$, i.e. the symbol $(a)$, in an English text body, then glue them together to form a new symbol $(ab)$ and replace 
occurrences of symbols $(a)(b)$ in your text body with $(ab)$. In this case you add the symbol 
$(ab)$ to your alphabet. The parsing depth of the alphabet rises from $n$ to $n+1$. 
If in this extended alphabet you already find another symbol $(solute)$, then in order to get the next larger alphabet you could apply the substitution $(ab)(solute)$ to $(absolute)$. The particular substitution transformation $(a)(b)\to (ab)$
implements a parsing rule, the structural shape of the parsing rule $[i\ j \to k]$, with $i$, $j$, and $k$ being variables for symbols, implements a parsing  rule template. 
We will call $[i\ j \to k]$ the elementary template. In fact one has to also specify how 
indices get selected to become unique and invertible. For a parsing rule to be invertible
we want to be able to perform the inverse substitution rule$ [k \to i\ j]=[i\ j \to k]^{-1}$, i.e.
to expand symbol $k$ into the subsequent occurrence of letters $ij$. For this to be possible 
only one parsing rule that produces a particular symbol $k$ may exist and $i$ and $j$
already have to exist to be selected. So if we identify $i$, $j$, and $k$ as the index of the letter,
and we have already have indices $1\cdots W$, then we can choose any $i,j\leq W$ and identify $k=W+1$. In this case we extended the base alphabet $\mathcal{A}_0$ containing $W$ symbols to 
 $\mathcal{A}_1$ containing $W+1$ symbols. In general we extend $\mathcal{A}_n$ containing $W+n$ symbols to $\mathcal{A}_{n+1}$ containing $W+n+1$ symbols.
\\
 
\textbf{Parsing rules and templates: an example}: 
Consider again English text all written solely in lower case letters so that our original alphabet 
consists of $W=27$ symbols, $1\equiv (a),\,2\equiv (b),\cdots,\,26\equiv (z),27={\rm SPACE}$.
We therefore start with $\mathcal{A}_n$ with $n=0$.
One might find the following 
sequence of parsing rules
$[(w)(o)\to(wo)]$, $[(wo)(r) \to (wor)]$ and $[(wor)(d)\to (word)]$ in our sequence of parsing rules linking the letter to the word level alphabet. Those rules would rewrite all subsequent occurrences of the letter w, o, r, and d in an English text body written in Latin letters with the symbol $(word)$ in 
the word level alphabet. The parsing rules we show here follow, what we call the 
\textit{elementary} parsing rule template $[i\ j\to k]$, with $i\equiv(\lambda_1)$ and 
$j\equiv(\lambda_2)$,
and $k\equiv (\lambda_1\lambda_2)$ being variables for symbols.
In this words Lempel-Ziv codecs \cite{LempelZiv} essentially rely on the elementary parsing rule template,
however, with the aim to compress (see above) and not primarily for efficiently extracting features.  
Taking the first parsing rule $[(w)(o)\to (wo)]$ that extends the original alphabet
identifies the new symbol $28=W+n+1\equiv (wo)$ and the parsing rule taken from the template 
$[i\ j\to k]$ reads
$[15\ 23\to 28]$, where $15\equiv (o)$ and $23\equiv (w)$. $28$ is the new symbol index in the extended alphabet $n=1$. Every new parsing on an alphabet $\mathcal{A}_n$ will yield a new symbol
with index $k=W+n+1$.

%There exist arbitrary complex parsing rule templates. 
%For simplicity's sake we will only consider the elementary parsing rule template in the main text and only consider process classes for which parsing templates other than the elementary template are irrelevant for feature extraction. See SI Text 3 for details on parsing rules templates, their relevance for estimating information production, and their connection to generative grammars, \cite{Chomsky1,Chomsky2}.

\textbf{More general parsing rules}:
However, in principle, depending on the process in question, one might also need to consider arbitrary complex templates. For instance, to capture clauses of the form: ``\textbf{ if} this \textbf{ then} that'', where "this" and "that" represents a text of variable length. The associated parsing rule template could read $i\ X\ j\ \to\ k[X]$ such that we can reversibly parse 
``(\textbf{ if})( this)(\textbf{ then})( that) $\to$ (\textbf{ if then})[( this)]( that)''. Note that we require additional parenthesis symbols $[$ and $]$ in order to make the parsing rule reversible.

Note also that we use parsing rules in the ``analytic mode'',  $[i\ j \to k]$, when we extend an alphabet
by one symbol. We use a parsing rule in the ``generative mode'' $[k \to i\ j]$ if we expand 
symbols $k=W+n+1$ to map messages over $\mathcal{A}_{n+1}$ to messages over $\mathcal{A}_n$.

Parsing rules have been studied intensively in the field of theoretical Linguistics as methods of generating text in terms of what are called generative grammars, \cite{Chomsky1}. 
To give a simple example: the parsing rules $X\to aX$ and $X\to bX$, with terminal $X\to \epsilon$, where $\epsilon$ is the \textit{empty symbol}, can be used to write any sequence of $a$'s
 and $b$'s; e.g.: $X\to aX\to aaX\to aabX\to\cdots\to aababbbaabaX \to aababbbaaba$.
 The two rules $X\to aX$ and $X\to Xb$ with the same terminal rule, on the other hand, would only produce sequences of the form $a^mb^n$, i.e. $X\to aX\to aaX\to aaXb\to\cdots\to aaaaaaXbbbb \to aaaaaabbbb$. It is a major achievement in this line of theoretical work that generative grammars can be classified in four hierarchically inclusive classes, \cite{Chomsky1,Chomsky2}, of grammars being so called regular grammars, and at the top level one finds everything you could write as a computer program on a universal touring machine with unbounded memory (recursively enumerable languages).

\subsection{SI Text 4: Information production and Kolmogorov complexity}

Kolmogorov complexity and information production are two closely related concepts. The one, Kolmogorov complexity, \cite{Kolmogorov,Solomonov,Chaitin}, basically refers to the length of the shortest program (in some universal computing language, that in fact can be though of to be generated by a particular generative grammar and its associated parsing rules set to generative mode. Think for instance, the program of a standard random number generator $y$ as it is implemented in many computing Languages, or as they can be found in 
\textit{Numerical recipes in C. The art of scientific computing} (Press, William H et al, Cambridge University Press 1986, 1992). Those codes are relatively short, i.e. their Kolmogorov complexity 
is finite, and if run, produce pseudo random numbers.
The entropy of the pseudo random number sequence, being deterministic and periodic 
with an extremely long period, is actually vanishing, if we had sufficient data to infer the periodicity of the signal.
However, those numbers in general are astronomically large and although theoretically one could observe the periodic structure, practically this is not possible, i.e. the deterministic numbers of sequences $x$ produced by $y$ can hardly be distinguished from actual random number sequences by statistical statistical test.
So if you know the generator $y$ and the current random seed $s$ you can perfectly predict the next number the generator emits and how the seed $s$ updates, i.e. you can predict $(y,s)$ translated into a number $x(t)$ the generator emits at the $t$'h step.
That is, in order to find the true information production of the process, we would have to
reconstruct parsing rules that in the end could transform data back into the code of the random number generator and its initial random seed in order to do so. In this case data $x$, in the infinite size limit would always be transformed back into a finite length $L(Y)$ which essentially corresponds to the code of the random number generator $Y\equiv y$ characterized by the pair $(y,s)$, $y$ being the code and $s$ the seed value. 
That is, we understand $X=\pi Y$ to be the process we observe, which emits the data $x$, and $\pi$ describes the hardware that translates $Y$ into $X$. As a consequence, the information production, up to possibly a constant, is given by  $I(X)=\lim_{t\to\infty} L(Y)/t=0$.
In other words, if we have a consistent method to extract structure from data, such that asymptotically 
$I(x)\to I(X)$, then the Kolmogorov complexity $L(y)\sim t I(x)$ is essentially the minimal description length of the data. In general we can expect that asymptotically $L(y)\propto t^{\alpha}$ for some exponent
$0\leq \alpha\leq 1$, measuring ``how deterministic a process is'', $\alpha=0$ being deterministic
programs (including pseudo random number generators) and $\alpha=1$, random processes
with a finite information production. 

It is however more than doubtful that it is possible to reconstruct
the parsing rules of arbitrary complex generative grammars purely from statistical analysis of the data they generate alone. This issue touches Chaitin's incompleteness theorem,
\cite{Chaitin}, which essentially states that above a certain string complexity it is no longer possible to decide whether a string is complex or not, i.e. whether it is still compressible or not. 
Intuitively we would assume that for our random generator example it is probably impossible to reconstruct 
some version of the random number generator program from the pseydo random numbers it generates, at least if you do not a priorly know that those numbers have been produced by a random number generator.

\subsection{SI Text 5: Detailed algebraic steps for Eq. (\ref{eq:MarkovSSREquality})}

Since we are dealing with a Markov process we can compute the information production also through the conditional entropy, that
as we will see, leads to the same maximum configuration of marginal distributions of state visits. We focus on the joint probabilities $p(i,j)$ to observe $i$ following $j$. From $p(i,j)$ one gets the marginal distribution, $p_i$, by marginalization. 
The question is how many sequences $x$ exist when we observe a  joint histogram $h(i,j)=Np(i,j)$. 
In other words, what is the multiplicity, $M$, of possible sequences $x$ and its associated reduced Boltzmann entropy $S$ and cross entropy $S^{\rm cross}$ if we the underlying Markov process is characterized by the transition probabilities, $q(i|j)$. 
It is not difficult to see that $S=\log(M)/N$, is given by the conditional entropy
\begin{equation}
S_{\rm cond}[p]\ =\ -\sum_{i,j=1}^W p(i,j)\ \log \frac{p(i,j)}{\sum_{m=1}^W p(m,j)}   \, .
\label{eq:MarkovEnt}
\end{equation}
%Note that in order to write $S$ as a functional of the joint distributions $p(i,j)$ we have avoided replacing $\sum_{m=1}^W p(m,j)=p_j$ and $p(i,j)=p(i|j)p_j$. This is necessary, since otherwise we are dealing with a functional that depends simultaneously on marginal and conditional probabilities, which not only makes the functional structurally more complicated, we would also have to add additional constraints that describe the relation between marginal and conditional probabilities. 
Similarly, the cross entropy of the Markov process is found
\begin{equation}
S^{\rm cross}_{\rm cond}(p|q)\ =\ -\sum_{i,j=1}^W p(i,j)\, \log\, q(i|j) \, .
\label{eq:MarkovCrossEnt}
\end{equation}
Note, that  $p(i,j)$ is the joint frequency distribution of transitions $j\to i$ in a sequence $x$, and $q(i|j)$ 
is the conditional probability distribution defining the  Markov process.
Maximizing $S_{\rm cond}-S^{\rm  cross}_{\rm  Markov}$ (the negative conditional information divergence) with respect to the joint distribution 
$p$ under the constraint, $\sum_{i,j} p(i,j)=1$, yields the expected result for the maximiser
\begin{equation}
p(i|j)\ =\ q(i|j) \, ,
\label{eq:MarkovSolution}
\end{equation}
and the transition probabilities $q(i|j)$ of the Markov process can be estimated asymptotically (large $N$) by the observed empirical conditional probabilities $p(i,j)/p_j$. If we do this for the 
slowly driven SSR process, then $p(i|j)=q(i|j)$ and the maximizing marginal distribution $p_i$ is obtained by solving the eigenvector equation $p_i=\sum_j q(i|j)p_j$. Since the conditional entropy, $S_{\rm cond}$, from Eq. (\ref{eq:MarkovEnt}) is the log of the multiplicity of sequences compatible with the empirical joint distribution
$p(i,j)$, its value should equal the value of $S_{\rm SSR}$ for the marginal distribution $p_i$ from Eq. (\ref{eq:SSRfin}), which is the corresponding maximizer. 
Denoting the maximizer of the joint distribution by $p^{(2)}$ and the 
one of the marginal distribution by $p^{(1)}$, and using  
$\frac{q_j}{Q_{j-1}Q_j}\ =\ \frac{1}{Q_{j-1}}-\frac{1}{Q_{j}}$, within a few algebraic steps (see below) we see that indeed
\begin{equation}
S_{\rm SSR}(p^{(1)})\ =\ S_{\rm cond}(p^{(2)})
\label{eq:MarkovSSREquality}
\end{equation}
holds identically for
% {\re $q(i|j)$ from Eq. (\ref{eq:SSRtransprob}) CHECK ORDER OF i and j}, 
$q(i|j)$ from Eq. (\ref{eq:SSRtransprob}), 
for all choices of weights $q_j\ \geq\ 0$ and $\sum_i q_i\ =\ 1$.
%{\bl For details, see SI Text 4.}
This implies that entropy in the context of complex processes can be approached exactly 
by means of \textit{information theory} and that the existence of generalized entropies is a 
consequence of the complex, non-i.i.d. structure of the underlying systems and processes. 
\\

{\bf Few algebraic steps}:
We first of all note that in maximum configuration the marginal distribution of the SSR process is given by 
$p^{(1)}_i=p^{(1)}_1q_i/Q_i$, where $p^{(1)}_1=1/Z$ acts as a normalization constant, $q_i$ is the weight distribution and $Q_i=\sum_{j=1}^i q_j$ is the cumulative weight distribution of the SSR process; $i=1,2,\cdots,W$. 

Using $p^{2}(i,j)=q(i|j)p^{(1)}_j$ and Eq. (\ref{eq:SSRtransprob}), we can compute
\begin{equation}
%\begin{array}{lcll}
%Z\ S_{\rm cond}(p^{(2)}) &=& -Z\sum_{i,j=1}^W p^{(1)}_i q(i|j)\ \log\left(  q(i|j) \right) &\\ 
%&=& -\sum_{i<j}  \frac{q_j}{Q_j} \frac{q_i}{Q_{j-1}}\ \log\left(  q_i \right)\qquad & {\rm (A)}\\
%&& +\sum_{i<j} \frac{q_j}{Q_j} \frac{q_i}{Q_{j-1}}\ \log\left(  Q_{j-1} \right) & {\rm (B)}\\
%&& + H(q) & {\rm (C)}
%\end{array}
\begin{array}{lcl}
S_{\rm cond}(p^{(2)}) &=& -\sum_{i,j=1}^W p^{(1)}_i q(i|j)\ \log\left(  q(i|j) \right)\\ 
&=& (A+B+C)/Z\,, \\ 
A&=& -\sum_{i<j}  \frac{q_j}{Q_j} \frac{q_i}{Q_{j-1}}\ \log\left(  q_i \right)\,,\\
B&=& \sum_{i<j} \frac{q_j}{Q_j} \frac{q_i}{Q_{j-1}}\ \log\left(  Q_{j-1} \right)\,,\\
C&=& H(q)\,.
\end{array}
\label{eq:SI:MarkovEnt}
\end{equation}
For term (A) we can use that 
\begin{equation}
\frac{q_i}{Q_iQ_{i-1}}=\frac{1}{Q_{i-1}}-\frac{1}{Q_{i}}
\end{equation} 
and $Q_W=1$ to compute $A$ to be given by  
\begin{equation}
\begin{array}{l}
%\begin{array}{lcl}
%A&=&\sum_{i=1}^{W-1}\sum_{j=i+1}^{W}\left( \frac{1}{Q_{j-1}}-\frac{1}{Q_{j}}\right)q_i\log q_i&=& \\
%&=&\sum_{i=1}^{W-1}\left( \frac{1}{Q_{i}}-1\right)q_i\log q_i\\
%&=&\sum_{i=1}^{W-1}\frac{1}{Q_{i}}q_i\log q_i\ -\ H(q)
%%%%%%%%%%%%%%%
\sum_{i=1}^{W-1}\sum_{j=i+1}^{W}\left( \frac{1}{Q_{j-1}}-\frac{1}{Q_{j}}\right)q_i\log q_i= \\
=\sum_{i=1}^{W-1}\left( \frac{1}{Q_{i}}-1\right)q_i\log q_i\\
=\sum_{i=1}^{W-1}\frac{1}{Q_{i}}q_i\log q_i\ -\ H(q)\,,
\end{array}
\end{equation}
Similarly, we get $B$;
\begin{small}
\begin{equation}
\begin{array}{l}
%\begin{array}{lcl}
%B&=&\sum_{i=1}^{W-1}\sum_{j=i+1}^{W}\left( \frac{1}{Q_{j-1}}-\frac{1}{Q_{j}}\right)q_i\log Q_{j-1}\\
%&=&\sum_{i=1}^{W-1}\sum_{j=i+1}^{W}\left( \frac{\log Q_{j-1}}{Q_{j-1}}-\frac{\log Q_{j-1}}{Q_{j}}\right)q_i\\
%&=&\sum_{i=1}^{W-1}q_i\left( \frac{\log Q_i}{Q_i}-\sum_{j=i+1}^W  \frac{1}{Q_i}\log\left(1-\frac{q_i}{Q_i}\right)\sum_{i=1}^{W-1}\sum_{j=i+1}^{W}\left( \frac{1}{Q_{j-1}}-\frac{1}{Q_{j}}\right)q_i\log Q_{j-1}= \\
%&=&\sum_{i=1}^{W-1}q_i\frac{\log Q_i}{Q_i}\ -\ \sum_{i=2}^W  \left(1-\frac{q_i}{Q_i}\right)\log\left(1-\frac{q_i}{Q_i}\right)
%%%%%%%%%
\sum_{i=1}^{W-1}\sum_{j=i+1}^{W}\left( \frac{\log Q_{j-1}}{Q_{j-1}}-\frac{\log Q_{j-1}}{Q_{j}}\right)q_i=\\
=\sum_{i=1}^{W-1}q_i\left( \frac{\log Q_i}{Q_i}-\sum_{j=i+1}^W  \frac{1}{Q_j}\log\left(1-\frac{q_j}{Q_j}\right)\right)\\
=\sum_{i=1}^{W-1}q_i\frac{\log Q_i}{Q_i}-\sum_{i=2}^W  \left(1-\frac{q_i}{Q_i}\right)\log\left(1-\frac{q_i}{Q_i}\right)
\end{array}
\end{equation}
\end{small}
Inserting the terms $A$, $B$, and $(C)$ into Eq. (\ref{eq:SI:MarkovEnt}) one gets 
\begin{small}
\begin{equation}
\begin{array}{lcl}
Z\ S_{\rm cond}(p^{(2)}) &=& -\sum_{i=1}^{W-1} \frac{q_i}{Q_i}\log\left(\frac{q_i}{Q_i}\right) \\ 
&& -\sum_{i=2}^{W} \left(1-\frac{q_i}{Q_i}\right)\log\left(1-\frac{q_i}{Q_i}\right) \\ 
&& -q_W\log(q_W)\ . \\ 
\end{array}
\label{eq:SI:MarkovEnt2}
\end{equation}
\end{small}
Again, $p^{(1)}_i=p^{(1)}_1 q_i/Q_i$ and $p^{(1)}_1=1/Z$. Also $Q_W=1$ and $q_1=Q_1$. 
One obtains that
$S_{\rm cond}(p^{(2)})$ equals to 
\begin{small}
\begin{equation}
%\begin{array}{l}
-\sum_{i=2}^W\left[p^{(1)}_i\log\frac{p^{(1)}_i}{p^{(1)}_1}+(p^{(1)}_1-p^{(1)}_i) \log\left(1-\frac{p^{(1)}_i}{p^{(1)}_1}\right)\right]\,.
%\end{array}
\end{equation}
\end{small}

This however is exactly $S_{\rm SSR}(p^{(1)})$ and therefore it follows that 
$S_{\rm cond}(p^{(2)})=S_{\rm SSR}(p^{(1)})$ for all possible choices of weights $q$.

\subsection{SI Text 6: An algorithm for estimating MDL}

There exist various strategies that one could follow in composing algorithms that generate, if not minimal adjoint alphabets from data, then at least reasonably small adjoint alphabets with maps $\pi$, even if we only consider the elementary parsing rule template.
One possible strategy would be to look for the longest repeated sequences in data, which are very unlikely to be produced by chance, given the marginal frequency distribution of letters, and then work ones way downward 
by intersecting such sequences untill one has decomposed all such pattern into elementary parsing rules.   
Despite the fact that the described top down strategy can be expected to be far superior in case we are dealing with non Markovian processes, we only implemented a simple bottom up strategy for
extracting parsing rules drawn from the 
elementary parsing rule template $[i\ j\to k]$ on statistical grounds.
It performs admirably, despite many shortcomings. It does produce relatively small but not optimal alphabets
for processes such as SSR processes. It mistakes relatively simple non Markovian processes to be purely random, i.e. the depth of the statistical analysis the algorithm performs cannot distinguish such processes from i.i.d. processes.
However, it already extracts much more information from written text than we would estimate with the conditional entropy of letter transition frequencies. The algorithm works as follows.

(i) To get the initial condition of the parsing process, take your data and 
transform it into an index sequence $x$ of symbols $i=1,\cdots, W$ with length $t=|x|$, that 
represent the original alphabet $\mathcal{A}_0$. You set $W_0=W$ and $y_0=x$, 
measure the histograms $h^0_i=h(x_0)$ and set the length of the data to $t_0=t$,
compute the distribution function $f^0=h^0/t_0$, set the average complexity 
of the alphabet to $\bar\ell_0=1$ and the MDL per emitted symbol 
$L_0=k H(f^0_i/t_0)/\bar\ell_0$. Then allocate an empty list of parsing rules and the original alphabet
$\mathcal{A}_0$ and the associated length of the symbol in original letters $\ell_i=1$. 
We write symbolically $i\to (i)$ for this representation. For instance, 
if we deal with the Latin alphabet this could mean $1\to (a)$, $2\to (b)$, and so forth.

We are now ready to iterate the state at parsing level $n$, starting from $n=0$.

(ii) Suppose the parsing level is $n$. 
Make a list of all symbol pairs $i$ following $j$ in the data $y_n$. Note that this list cannot be longer
than $\min(W_n^2,t_n-1)$. Then iterate over all pairs in the list and tentatively apply the 
parsing rule $[i\ j\to W_n+1]$, i.e. we generate the symbol $(ij)$ from $(i)(j)$,
and count the number of (non overlapping) occurrences $h_{ij}$.
Without the need to actually generate the data $y_{n+1}$ for the particular parsing rule
one can compute several things, starting by the complexity of the new symbol 
$\ell_{W_n+1}=\ell_{i}+\ell_{j}$. Let us assume that $i\neq j$, i.e. the two symbols of the pair are differ so that pairs cannot have an overlap with themselves.
If they are the same a similar rule applies, only we have to count for possible overlaps, eg. if one finds
tripples $iii$ ($h_{ii}=1$), or quadruples $iiii$ ($h_{ii}=2$) and so forth.
We also can compute how many pairs we would expect to find if the data were already i.i.d.
and compute $h^*_{ij}=h^{n}_i f^{n}_j$ (the formula for symbol pairs with potential overlap is slightly different).
We therefore know that after the substitution we get $t_{n+1}=t_{n}-n_{ij}$,
$h^{n+1}(W_n+1)=h_{ij}$, $h^{n+1}(i)=h^{n}(i)-h_{ij}$ and also $h^{n+1}(j)=h^{n}(j)-h_{ij}$,
while for all other symbols $z$ one gets $h^{n+1}(z)=h^{n}(z)$.
From this we can compute $f^{n+1}$ and all values that define the new state.
In particular we can compute the new tentative $\bar\ell_{n+1}$, by computing $\sum_{z=1}^{W_n+1}\ell(z)f^{n+1}(z)$ and the new $L_{n+1}=k H(f^{n+1}_i/t_{n+1})/\bar\ell_{n+1}$.
We can do the same thing for the i.i.d. estimate for the expected number of pairs $h^*_{ij}$.
In particular we get an i.i.d. estimate for the new MDL  $L^*_{n+1}=k H(f^{*}_i/t^*)/\bar\ell^*$.
From this we compute two numbers for each pair. The first is the number of
standard deviations $h_{ij}$ differs from $h^*_{ij}$. Using $f_{ij}=h_{ij}/t_n$ and 
$f^*_{ij}=h^*_{ij}/t_n=f^{n}_i f^{n}_j$ one obtains:
\begin{equation}
n_{\rm std}=\frac{f_{ij}-f^*_{ij}}{\sqrt{f^*_{ij}(1-f^*_{ij})}}\,.
\end{equation}
If the pair $ij$ has already been screened at an earlier time, then we remember the maximum 
of the previous $n_{\rm std}$ and the current number. Note that in general $L^*_{n+1}>L_{n+1}$.
The second number is simply, $\Delta L=L^*_{n+1}-L_{n+1}\geq 0$, 
the amount of description length reduction we get, relative to what we would expect from an i.i.d. process. 

(iii) After we have computed those two numbers for any pair in the list we select a pair in the following way. We determine the maximal value $\Delta L^{\rm max}$ of $\Delta L$ over all pairs and the maximal values $n^{\rm max}_{\rm std}$ of $n_{\rm std}$. Then we look for the for all pairs with values of $\Delta L >\gamma \Delta L^{\rm max}$ and $n_{\rm std}>\gamma n^{\rm max}_{\rm std}$.
We found $\gamma=0.1$ to work well. Then we random sequentially pick any of those pairs until 
we find one such that $L_n/L_{n+1}-1>\epsilon$, where $\epsilon \gtrsim 0$ (e.g. $\epsilon=10^{-5}$).
If we find such a pair, then select the associated parsing rule and store it in the list of 
parsing rules together with the newly generated symbol $(ij)\equiv W_{n}+1$ 
and its complexity $\ell(W_{n}+1)=\ell(i)+\ell(j)$. 
Apply the parsing rule to the data $y_n$ to get $y_{n+1}$. Update all values,
$W_{n+1}=W_{n}+1$, $\bar\ell_{n+1}$, and all the histograms, and so forth. 
Then go back to (ii) and iterate until we cannot select any new pair any more 
that would yield $L_n/L_{n+1}-1>\epsilon$
and the algorithm stops.
\\

The algorithm is very simple and works well enough for a proof of the concept.
However it has several shortcomings we believe can be overcome by implementing a top to bottom strategy
and we will therefore not publish this simple bottom to top algorithm in its current form. 
However, it provides us with a starting point for investigating how parsing rules can be efficiently constructed from finite amounts of data and to explore practical and theoretical limits of statistical inference 
to distinguishing structure from noise.  One can also attack questions about whether or not the elementary parsing rule template is being the only relevant template for a particular process class and the theoretical means to infer the applicability of more complex parsing rule templates. Another line of questions will have to deal with processes where the original alphabet is already very large in its own right (e.g. images forming a motion picture) where error free encoding may prove an unrealistically hard demand.
However, we intend to discuss improved versions of the algorithm 
and its application to data at some other place.
\end{document}